\documentclass[conference]{IEEEtran}
\usepackage[spanish,mexico]{babel}
\usepackage[utf8]{inputenc}

\ifCLASSINFOpdf
\else
\fi
\begin{document}
%
\title{Efectividad de técnicas de prueba de software aplicadas por sujetos novicios de pregrado}

\author{\IEEEauthorblockN{Omar S. Gómez}
\IEEEauthorblockA{Prometeo Senescyt - E.S. Politécnica de Chimborazo\\
Riobamba, Chimborazo, Ecuador; \\
Universidad Autónoma de Yucatán\\
 Mérida, México, ogomez@espoch.edu.ec \\
}
\and
\IEEEauthorblockN{Raúl A. Aguilar}
\IEEEauthorblockA{Facultad de Matemáticas\\
Universidad Autónoma de Yucatán\\
Mérida, Yucatán, México\\
Email: avera@uady.mx}
\and
\IEEEauthorblockN{Juan P. Ucán\\}
\IEEEauthorblockA{Facultad de Matemáticas\\
Universidad Autónoma de Yucatán\\
Mérida, Yucatán, México\\
Email: juan.ucan@uady.mx}}


%


\maketitle

\begin{abstract}
El objetivo de este trabajo se centra en investigar los efectos que conlleva realizar experimentos en ingeniería de software (IS) empleando como sujetos experimentales a estudiantes de pregrado cursando su primer año de estudios de la carrera en ingeniería de software. De manera particular en este trabajo se investiga la efectividad medida en porcentaje de defectos observados y observables de las técnicas de prueba de software funcional (caja negra) y estructural (caja blanca). 

Con respecto a los defectos observados por los sujetos, ambas técnicas obtuvieron una efectividad del 4\%. Con respecto a los defectos observables por los casos de prueba, la técnica funcional es ligeramente superior (21\%) que la técnica estructural (16\%), aunque esta diferencia no es significativa. Se observa un nivel de inexperiencia considerable en los sujetos para aplicar las técnicas. Dado los hallazgos encontrados, se sugiere emplear sujetos de pregrado con un nivel mayor de experiencia.
\end{abstract}


%
\IEEEpeerreviewmaketitle

\section{Introducción}
La detección de defectos es una actividad crítica en la construcción de sistemas software. En general, la comunidad en ingeniería de software acepta que la detección de defectos debiera llevarse a cabo tanto con técnicas de inspección basadas en lectura de códigos así como por técnicas de prueba dinámicas como son las técnicas funcional y estructural, conocidas también como técnicas caja negra y caja blanca, respectivamente \cite{McConnell2004}. 

El trabajo aquí reportado es una extensión de trabajos previos \cite{Basili1987,Kamsties1995,Roper1997,Juristo2003} en donde se ha investigado, entre otros aspectos, la efectividad de varias técnicas de prueba de software. A diferencia de estos trabajos previos, en esta investigación se busca analizar la efectividad de las técnicas de caja negra y caja blanca aplicadas por sujetos novicios de pregrado, con el fin de realizar recomendaciones sobre el uso de este tipo de sujetos en experimentos controlados en ingeniería de software. El resto de este documento se estructura de la siguiente manera: en la sección \ref{sec:2} se describen las dos técnicas de prueba estudiadas. En la sección \ref{sec:3} se presenta el contexto del experimento. En la sección \ref{sec:4} se describe el análisis estadístico. Por último, en la sección \ref{sec:5} se presenta la discusión y conclusiones.

\section{Técnicas de prueba de software}
\label{sec:2}

\subsection{Técnica funcional o caja negra}
También conocida como prueba de comportamiento, este tipo de técnica se basa en la especificación del programa o componente a ser probado para elaborar los casos de prueba. El componente se ve como una ``caja negra'' cuyo comportamiento sólo puede ser determinado estudiando sus entradas y las salidas obtenidas a partir de ellas. No obstante, como el estudio de todas las posibles entradas y salidas de un programa es impracticable se selecciona un conjunto de ellas sobre las que se realizan las pruebas. Para seleccionar el conjunto de entradas y salidas sobre las que trabajar, hay que tener en cuenta que en todo programa existe un conjunto de entradas que causan un comportamiento erróneo en el sistema, y como consecuencia producen una serie de salidas que revelan la presencia de defectos. Entonces, dado que la prueba exhaustiva es imposible, el objetivo final es encontrar una serie de datos de entrada cuya probabilidad de pertenecer al conjunto de entradas que causan dicho comportamiento erróneo sea lo más alto posible \cite{Beizer1990,Myers2004}.

Para especificar los casos de prueba de caja negra existen distintos criterios, algunos de ellos son:
\begin{itemize}
\item \textbf{Particiones de clase de equivalencia.} Una clase de equivalencia representa un conjunto de estados válidos o no válidos para condiciones de entrada. Típicamente una condición de entrada es un valor numérico específico, un rango de valores, un conjunto de valores relacionados o una condición lógica. 

\item \textbf{Análisis de Valores Límite.} los errores tienden a darse más en los límites del campo de entrada que en el centro. El análisis de valores límite lleva a una elección de casos de prueba que ejerciten dichos valores. El análisis de valores límite (AVL) es una técnica de diseño de casos de prueba que completa a la partición equivalente. En lugar de seleccionar cualquier elemento de una clase de equivalencia, el AVL lleva a la elección de casos de prueba en los extremos de la clase. En lugar de centrarse solamente en las condiciones de entrada, el AVL obtiene casos de prueba también para el campo de salida.
\end{itemize}

\subsection{Técnica estructural o caja blanca}
A este tipo de técnica se le conoce también como técnica de caja transparente o de cristal. Esta técnica se centra en cómo diseñar los casos de prueba atendiendo al comportamiento interno y la estructura del programa. Se examina así la lógica interna del programa sin considerar los aspectos de rendimiento.

El objetivo de la técnica es diseñar casos de prueba para que se ejecuten, al menos una vez, todas las sentencias del programa, y todas las condiciones tanto en su vertiente verdadera como falsa \cite{Beizer1990,Myers2004}. 

Como se ha indicado, puede ser impracticable realizar una prueba exhaustiva de todos los caminos de un programa. Por ello se han definido distintos criterios de cobertura lógica, que permiten decidir qué sentencias o caminos se deben examinar con los casos de prueba. Estos criterios son:

\begin{itemize}
\item \textbf{Cobertura de sentencias.} Se escriben casos de prueba suficientes para que cada sentencia en el programa se ejecute, al menos, una vez.

\item \textbf{Cobertura de decisión.} Se escriben casos de prueba suficientes para que cada decisión en el programa se ejecute una vez con resultado verdadero y otra con el falso.

\item \textbf{Cobertura de condiciones.} Se escriben casos de prueba suficientes para que cada condición en una decisión tenga una vez resultado verdadero y otra falso.


\item \textbf{Cobertura de condición múltiple.} Se escriben casos de prueba suficientes para que todas las combinaciones posibles de resultados de cada condición se invoquen al menos una vez. 

\item \textbf{Cobertura de caminos.} Se escriben casos de prueba suficientes para que se ejecuten todos los caminos de un programa. Entendiendo camino como una secuencia de sentencias encadenadas desde la entrada del programa hasta su salida. 
\end{itemize}

\section{Contexto del experimento}
\label{sec:3}

Tomando como referencia el proceso general de experimentación en ingeniería de software \cite{Gomez2013c} a continuación se detalla el experimento aquí reportado. 

\subsection{Definición}
Las hipótesis de trabajo para este experimento se definen de la siguiente manera: Existe al menos una técnica de prueba de software (técnica funcional por particionamiento de clases de equivalencia [PCE], o estructural por cobertura de decisiones [Cob. Dec.]) tal que ésta difiere del resto con respecto a su efectividad.

$H_{0}$: efectividad en $PCE=Cob. Dec.$

$H_{1}$: efectividad en $PCE \ne Cob. Dec.$

En este contexto, la efectividad es medida con dos métricas: defectos observados por los sujetos y defectos observables por los casos de prueba especificados por los sujetos. 

Los defectos observados indican el porcentaje de defectos que el sujeto es capaz de observar tras aplicar alguna de las técnicas de prueba de software.

Por otra parte, los defectos observables indican el porcentaje de defectos que son revelados por los casos de prueba especificados por los sujetos. 

Cabe señalar que el porcentaje de defectos observados puede ser distinto que el porcentaje de defectos observables. Por ejemplo, un sujeto quien especifica varios casos de prueba, que en conjunto revelan tres de seis posibles defectos, es capaz de observar sólo un defecto de los tres revelados por sus casos de prueba. En este caso el porcentaje de defectos observados es de 17\%, mientras que el porcentaje de defectos observables es de 50\%.

\subsection{Diseño}
Con el fin de estudiar posibles interacciones entre los efectos de la técnica y del programa instrumentado, se empleó un diseño factorial con dos factores y dos niveles. El primer factor representa la técnica de prueba y contiene dos niveles correspondientes con cada tipo de técnica: funcional y estructural. El segundo factor representa el programa instrumentado, en este caso se emplearon dos programas donde cada programa corresponde con un nivel de este segundo factor. En total se tienen cuatro posibles combinaciones entre los niveles de estos dos factores. 

Las combinaciones factor-nivel en el experimento son asignadas de forma aleatoria a las unidades experimentales, en este caso, los sujetos. Este tipo de diseño permite el estudio del efecto de cada factor (técnica y programa) sobre la variable respuesta (efectividad), así como el efecto de las interacciones entre factores sobre dicha variable (técnica-programa). En la Tabla \ref{tab:1} se muestra la estructura de este diseño así como las combinaciones que se asignaron de forma aleatoria a los sujetos.

\begin{table}[!h]
\renewcommand{\arraystretch}{1.3}
\caption{Diseño factorial empleado.}
\label{tab:1}
\centering
\begin{tabular}{|r||r||r|}
\hline
 Programa/Técnica& Funcional & Estructural \\
\hline
 cmdline & funcional, cmdline & estructural, cmdline \\
\hline
 ntree   & funcional, ntree 	& estructural, ntree \\
\hline
\end{tabular}
\end{table}

\subsection{Ejecución}
El experimento aquí reportado se realizó a inicios de marzo de 2013 en la asignatura de programación. Esta asignatura se imparte en el segundo semestre de la Licenciatura en Ingeniería de Software de la Facultad de Matemáticas de la Universidad Autónoma de Yucatán (FMat-UADY). Los alumnos participantes como sujetos en este experimento cursaron durante el semestre anterior la asignatura de fundamentos de programación.

Semanas previas al experimento los alumnos inscritos al curso de programación recibieron entrenamiento sobre el funcionamiento y aplicación de las técnicas de prueba de software. Una vez explicado el funcionamiento de las técnicas, los sujetos aplicaron cada técnica a un ejercicio de programación instrumentado con algunos defectos.

Ya finalizado el entrenamiento y en una sesión independiente se realizó el experimento. Se asignó de manera aleatoria las cuatro combinaciones de tratamientos descritas en la Tabla \ref{tab:1} a los sujetos. En total participaron 31 sujetos. Los sujetos usaron una herramienta web para registrar información referente al diseño de sus casos de prueba y defectos observados de acuerdo a la técnica y programa asignado.

Los programas instrumentados que se usaron en este experimento se escribieron en el lenguaje de programación C. En cada programa se inyectaron seis defectos con el fin de evaluar la efectividad de las dos técnicas de prueba. Los defectos se tomaron como referencia de la clasificación propuesta en \cite{Basili1984}. Cada programa tiene una longitud aproximada de 250 LOC.

A continuación se detalla de manera general el procedimiento de aplicación de ambas técnicas. Con respecto a la aplicación de la prueba funcional, el sujeto recibe la especificación del programa instrumentado sin tener acceso al código fuente. Con la especificación el sujeto define clases de equivalencia válidas y no válidas. A continuación se construyen casos de prueba usando como referencia las clases de equivalencia. Una vez especificados los casos de prueba el sujeto tiene acceso de ejecución al programa instrumentado y ejecuta sus casos de prueba. Esta actividad concluye cuando el sujeto registra las salidas observadas de los casos de prueba ejecutados. La siguiente actividad consiste en volver a revisar la especificación con el fin de identificar posibles defectos revelados en las salidas de los casos de prueba especificados. La aplicación de esta técnica concluye una vez que el sujeto registra los defectos observados.

Con respecto a las pruebas estructurales, el sujeto recibe el código fuente impreso pero no la especificación. A partir del código fuente el sujeto construye casos de prueba intentando cubrir el 100\% de las ramificaciones de flujo que pudiera alcanzar el código del programa. Una vez especificados los casos de prueba el sujeto usa una versión instrumentada del programa para ejecutar sus casos de prueba. Después de ejecutar los casos de prueba, el sujeto registra las salidas observadas. Posteriormente el sujeto accede a la especificación del programa con el fin de identificar posibles defectos en las salidas observadas. La aplicación de esta técnica finaliza con el registro de los defectos observados.

\section{Análisis}
\label{sec:4}
La ecuación (\ref{eq:1}) describe el modelo estadístico empleado dado el diseño factorial descrito en la sección \ref{sec:3}.

\begin{equation}
\label{eq:1}
y_{ijk}=\mu+\alpha_i+\beta_j+(\alpha\beta)_{ij}+\epsilon_{ijk}
\end{equation}

Donde $\mu$ es el promedio general, $\alpha_{i}$ representa el efecto de la técnica $i$, $\beta_{j}$ representa el efecto del programa $j$, $(\alpha\beta)_{ij}$ es el efecto de la interacción entre los tratamientos $i$ y $j$, $k$ representa las réplicas en cada combinación de tratamientos, y $\epsilon$ es el error aleatorio que asume una distribución normal de promedio $0$ y varianza $\sigma^2$. Este modelo emplea el análisis de varianza (ANOVA) para evaluar los diferentes componentes del modelo descrito como son: los tratamientos, sus interacciones así como el error aleatorio \cite{Juristo2001,Kuehl2000}.

El ANOVA proporciona una prueba estadística para determinar si los promedios de varios grupos de datos son todos iguales. La hipótesis nula asume que todos los grupos son simplemente muestras aleatorias de la misma población. Este supuesto implica que todos los tratamientos tienen el mismo efecto (quizás ninguno). Rechazar la hipótesis nula implica que los distintos tratamientos producen un efecto diferente. 

Antes de obtener cualquier conclusión sobre el modelo estadístico se deben evaluar los siguientes supuestos: 1) las observaciones, en este caso las métricas recabadas sean independientes (independencia); 2) la varianza sea la misma para todas las observaciones (homocedasticidad); las observaciones en cada grupo de tratamientos tengan una distribución normal (normalidad).

El primer supuesto se cumple por el principio de aleatoriedad usado en este diseño experimental, donde todas las mediciones de una muestra no están relacionadas con aquellas pertenecientes a otra muestra. Los otros dos supuestos se evalúan a través de los residuos estimados (i.e. error aleatorio, $\epsilon_{ijk}$) \cite{Box1978,Kuehl2000}. 

El supuesto de homocedasticidad puede evaluarse a través de la prueba estadística de Levene \cite{Levene1960} donde un valor $p$ significativo indica desigualdad de varianzas por lo que el supuesto de homocedasticidad es violado. 

Por otra parte el supuesto de normalidad puede evaluarse con la prueba estadística de Kolmogorov-Smirnov \cite{Kolmogorov1933,Smirnov1948}, donde un valor $p$ significativo indica una diferencia entre la distribución del estadístico (en este caso los residuos estandarizados del conjunto de observaciones) y la distribución poblacional (en este caso una distribución normal). Esta diferencia significativa sugiere una violación al supuesto de normalidad.

\subsection{Análisis con respecto a defectos observados}
En la Tabla \ref{tab:2} se presenta el ANOVA con respecto a los defectos observados por los sujetos tras aplicar las técnicas.

\begin{table}[!h]
\renewcommand{\arraystretch}{1.3}
\caption{ANOVA con respecto a defectos observados.}
\label{tab:2}
\centering
\begin{tabular}{|l||r||r||r||r||r|}
\hline
Component & Df & Sum Sq & Mean Sq & F value & p-value\\
\hline
técnica 			& 1 & 0.6 	& 0.60 	& 0.011 & 0.916 \\
programa 			& 1 & 128.5 	& 128.49 	& 2.424 & 0.131 \\
técnica:programa & 1 & 89.0 	& 88.97 	& 1.678 & 0.206 \\
residuals			& 27& 1431.3 & 53.01 	& 		 & \\
\hline
\end{tabular}
\end{table}

Como se observa en en la Tabla \ref{tab:2}, los factores principales (técnica y programa) así como la interacción técnica-programa no muestran diferencias significativas, es decir, ambas técnicas muestran una efectividad similar.  

Con respecto al supuesto de normalidad, la prueba de Kolmogorov-Smirnov \cite{Kolmogorov1933,Smirnov1948} arroja un valor $p = 0.04719$ indicando falta de normalidad en los residuos estandarizados. Esta falta de normalidad se presenta debido a que la mayoría de sujetos no observaron defectos en los programas. Referente al supuesto de homocedasticidad, la prueba de Levene \cite{Levene1960} arroja un valor $p=0.2727$ sugiriendo la aceptación de la hipótesis nula a favor de la igualdad de varianzas. 

En la Tabla \ref{tab:3} se muestran los análisis descriptivos de los factores principales técnica y programa.

\begin{table}[!h]
\renewcommand{\arraystretch}{1.3}
\caption{Análisis descriptivos con respecto a defectos observados.}
\label{tab:3}
\centering
\begin{tabular}{|l||r||r||r||r||r|}
\hline
					 	& n & mean & sd & min & max \\
\hline
técnica funcional 	& 16 & 4.17\% & 7.46 & 0\% & 16.67\% \\
técnica estructural & 15 & 4.45\% & 7.63 & 0\% & 16.67\% \\
programa cmdline 	& 16 & 6.25\% & 8.34 & 0\% & 16.67\% \\
programa ntree		& 15 & 2.22\% & 5.87 & 0\% & 16.67\% \\
\hline
\end{tabular}
\end{table}

\subsection{Análisis con respecto a defectos observables}
En la Tabla \ref{tab:4} se presenta el ANOVA con respecto a los defectos observables por los casos de prueba especificados por los sujetos. 

\begin{table}[!h]
\renewcommand{\arraystretch}{1.3}
\caption{ANOVA con respecto a defectos observables por los casos de prueba.}
\label{tab:4}
\centering
\begin{tabular}{|l||r||r||r||r||r|}
\hline
Component & Df & Sum Sq & Mean Sq & F value & p-value\\
\hline
técnica 			& 1 & 216 & 215.8 	& 0.791 & 0.382 \\
programa 			& 1 & 2 	& 1.7 		& 0.006 & 0.938 \\
técnica:programa & 1 & 389 & 389.2 	& 1.426 & 0.243 \\
residuals			& 27& 7367 & 272.9 	& 		 & \\
\hline
\end{tabular}
\end{table}

De acuerdo a los resultados de la Tabla \ref{tab:4}, los factores principales (técnica y programa) así como la interacción técnica-programa no muestran diferencias significativas, es decir, ambas técnicas muestran una efectividad similar.  

Con respecto al supuesto de normalidad, la prueba de Kolmogorov-Smirnov \cite{Kolmogorov1933,Smirnov1948} arroja un valor $p = 0.5312$ indicando la aceptación de la hipótesis nula a favor de la normalidad. Referente al supuesto de homocedasticidad, la prueba de Levene \cite{Levene1960} arroja un valor $p=0.2562$ sugiriendo la aceptación de la hipótesis nula a favor de la igualdad de varianzas. 

En la Tabla \ref{tab:5} se muestran los análisis descriptivos de los factores principales técnica y programa con respecto a esta métrica.

\begin{table}[!h]
\renewcommand{\arraystretch}{1.3}
\caption{Análisis descriptivos con respecto a defectos observables.}
\label{tab:5}
\centering
\begin{tabular}{|l||r||r||r||r||r|}
\hline
					 	& n & mean & sd & min & max \\
\hline
técnica funcional 	& 16 & 20.83\% & 12.91 & 0\% & 50\% \\
técnica estructural & 15 & 15.56\% & 19.38 & 0\% & 50\% \\
programa cmdline 	& 16 & 18.75\% & 15.96 & 0\% & 50\% \\
programa ntree		& 15 & 17.78\% & 17.21 & 0\% & 50\% \\
\hline
\end{tabular}
\end{table}

\section{Discusión y conclusiones}
\label{sec:5}

En la Tabla \ref{tab:6} se muestran resultados de trabajos previos que han analizado la efectividad de las técnicas aquí estudiadas. La efectividad está representada como el porcentaje de defectos observados por los sujetos tras aplicar alguna de las técnicas. 

\begin{table}[!h]
\renewcommand{\arraystretch}{1.3}
\caption{Defectos observados en trabajos previos.}
\label{tab:6}
\centering
\begin{tabular}{|l||c||c|}
\hline
	Experimento & Téc. Funcional & Téc. Estructural \\
\hline
Kamsties y Lott, 1994 \cite{Kamsties1995}& 51\% & 52\% \\
Kamsties y Lott, 1995 \cite{Kamsties1995}& 62\% & 56\% \\
Roper et al., 1997 \cite{Roper1997}& 55\% & 57\% \\
Juristo y Vegas, 2003 \cite{Juristo2003} & 62\% & 58\% \\
Experimento aquí reportado & 4.2\% & 4.5\% \\
\hline
\end{tabular}
\end{table}

Como se observa en la Tabla \ref{tab:6}, los resultados aquí reportados muestran un porcentaje inferior en la efectividad de las técnicas. A diferencia del resto de trabajos previos, en este experimento se emplearon estudiantes de pregrado cursando su primer año de la carrera, mientras que en los trabajos previos se emplearon estudiantes cursando su tercer y cuarto año de la carrera.

Con respecto a los defectos observables por los casos de prueba, se obtuvieron porcentajes mayores que los defectos observados. Por ejemplo, en el caso de la técnica funcional, los casos diseñados por los sujetos en promedio revelaron el 21\% de los defectos mientras que para la técnica estructural el 16\%. Los porcentajes de defectos observables arrojados en este experimento sugieren una falta de experiencia por parte de los sujetos para aplicar las técnicas aquí estudiadas. 

De manera general, los sujetos no fueron capaces de observar los defectos que revelaron sus casos de prueba. Se observa también que en el programa cmdline se identificaron el mayor número de defectos. Los resultados aquí reportados pueden servir como referente para que otros investigadores tengan en cuenta el nivel de experiencia de los sujetos al momento de realizar experimentos en esta disciplina.

\bibliographystyle{IEEEtran}
\bibliography{IEEEabrv,bibliography}

\begin{thebibliography}{10}
\providecommand{\url}[1]{#1}
\csname url@samestyle\endcsname
\providecommand{\newblock}{\relax}
\providecommand{\bibinfo}[2]{#2}
\providecommand{\BIBentrySTDinterwordspacing}{\spaceskip=0pt\relax}
\providecommand{\BIBentryALTinterwordstretchfactor}{4}
\providecommand{\BIBentryALTinterwordspacing}{\spaceskip=\fontdimen2\font plus
\BIBentryALTinterwordstretchfactor\fontdimen3\font minus
  \fontdimen4\font\relax}
\providecommand{\BIBforeignlanguage}[2]{{%
\expandafter\ifx\csname l@#1\endcsname\relax
\typeout{** WARNING: IEEEtran.bst: No hyphenation pattern has been}%
\typeout{** loaded for the language `#1'. Using the pattern for}%
\typeout{** the default language instead.}%
\else
\language=\csname l@#1\endcsname
\fi
#2}}
\providecommand{\BIBdecl}{\relax}
\BIBdecl

\bibitem{McConnell2004}
S.~McConnell, \emph{Code Complete}, 2nd~ed.\hskip 1em plus 0.5em minus
  0.4em\relax Redmond, WA, USA: Microsoft Press, 2004.

\bibitem{Basili1987}
V.~Basili and R.~Selby, ``Comparing the effectiveness of software testing
  strategies,'' \emph{IEEE Trans. Softw. Eng.}, vol.~13, no.~12, pp.
  1278--1296, 1987.

\bibitem{Kamsties1995}
E.~Kamsties and C.~M. Lott, ``An empirical evaluation of three defect-detection
  techniques,'' in \emph{Proceedings of the 5th European Software Engineering
  Conference}.\hskip 1em plus 0.5em minus 0.4em\relax London, UK:
  Springer-Verlag, 1995, pp. 362--383.

\bibitem{Roper1997}
M.~Roper, M.~Wood, and J.~Miller, ``An empirical evaluation of defect detection
  techniques,'' \emph{Information and Software Technology}, vol.~39, no.~11,
  pp. 763--775, 1997.

\bibitem{Juristo2003}
N.~Juristo and S.~Vegas, ``Functional testing, structural testing and code
  reading: What fault type do they each detect?'' in \emph{Empirical Methods
  and Studies in Software Engineering}, ser. Lecture Notes in Computer Science,
  R.~Conradi and A.~Wang, Eds.\hskip 1em plus 0.5em minus 0.4em\relax Springer
  Berlin / Heidelberg, 2003, vol. 2765, pp. 208--232.

\bibitem{Beizer1990}
B.~Beizer, \emph{Software testing techniques (2nd ed.)}.\hskip 1em plus 0.5em
  minus 0.4em\relax New York, NY, USA: Van Nostrand Reinhold Co., 1990.

\bibitem{Myers2004}
G.~J. Myers and C.~Sandler, \emph{The Art of Software Testing}.\hskip 1em plus
  0.5em minus 0.4em\relax John Wiley \& Sons, 2004.

\bibitem{Gomez2013c}
O.~S. Gómez, J.~P. Ucán, and G.~E. Gómez, ``Aplicación del proceso de
  experimentación a la ingeniería de software,'' \emph{Abstraction \&
  Application}, vol.~8, pp. 26--37, 2013.

\bibitem{Basili1984}
V.~Basili and B.~Perricone, ``Software errors and complexity: an empirical
  investigation,'' \emph{Commun. ACM}, vol.~27, no.~1, pp. 42--52, 1984.

\bibitem{Juristo2001}
N.~Juristo and A.~M. Moreno, \emph{Basics of Software Engineering
  Experimentation}.\hskip 1em plus 0.5em minus 0.4em\relax Kluwer Academic
  Publishers, 2001.

\bibitem{Kuehl2000}
R.~Kuehl, \emph{Design of Experiments: Statistical Principles of Research
  Design and Analysis}, 2nd~ed.\hskip 1em plus 0.5em minus 0.4em\relax
  California, USA.: Duxbury Thomson Learning, 2000.

\bibitem{Box1978}
G.~E.~P. Box, W.~G. Hunter, J.~S. Hunter, and W.~G. Hunter, \emph{Statistics
  for Experimenters: An Introduction to Design, Data Analysis, and Model
  Building}.\hskip 1em plus 0.5em minus 0.4em\relax John Wiley \& Sons, June
  1978.

\bibitem{Levene1960}
H.~Levene, ``Robust tests for equality of variances,'' in \emph{Contributions
  to probability and statistics}, I.~Olkin, Ed.\hskip 1em plus 0.5em minus
  0.4em\relax Palo Alto, CA: Stanford Univ. Press., 1960.

\bibitem{Kolmogorov1933}
A.~N. Kolmogorov, ``Sulla determinazione empirica di una legge di
  distribuzione,'' \emph{Giornale dell'Istituto Italiano degli Attuari},
  vol.~4, pp. 83--91, 1933.

\bibitem{Smirnov1948}
N.~V. Smirnov, ``Table for estimating the goodness of fit of empirical
  distributions,'' \emph{Ann. Math. Stat.}, vol.~19, pp. 279--281, 1948.

\end{thebibliography}
%



\end{document}